\documentclass[preprint]{aastex}

\newbox\grsign \setbox\grsign=\hbox{$>$} 
\newdimen\grdimen \grdimen=\ht\grsign
\newbox\laxbox \newbox\gaxbox
\setbox\gaxbox=\hbox{\raise.5ex\hbox{$>$}\llap
     {\lower.5ex\hbox{$\sim$}}}\ht1=\grdimen\dp1=0pt
\setbox\laxbox=\hbox{\raise.5ex\hbox{$<$}\llap
     {\lower.5ex\hbox{$\sim$}}}\ht2=\grdimen\dp2=0pt

\def\lax{\mathrel{\copy\laxbox}}

\begin{document}

\title{
Interstellar X-ray Absorption Spectroscopy of Oxygen, Neon, and Iron
with the Chandra LETGS Spectrum of X0614+091.
}

\author{
Frits Paerels\altaffilmark{1,2}, 
A. C. Brinkman\altaffilmark{2},
R. L. J. van der Meer\altaffilmark{2},
J. S. Kaastra\altaffilmark{2},
E. Kuulkers\altaffilmark{2,6},
A. J. F. den Boggende\altaffilmark{2},
P. Predehl\altaffilmark{3},
Jeremy J. Drake\altaffilmark{4}, 
Steven M. Kahn\altaffilmark{1}, 
Daniel W. Savin\altaffilmark{1}, and
Brendan M. McLaughlin\altaffilmark{4,5}
}
\altaffiltext{1}{Columbia Astrophysics Laboratory,
Columbia University, 538 W. 120th St., New York, NY 10027, USA} 

\altaffiltext{2}{SRON Laboratory for Space Research,
Sorbonnelaan 2, 3584 CA Utrecht, the Netherlands}

\altaffiltext{3}{Max Planck Institut f\"ur Extraterrestrische
Physik, Postfach 1503, D-85740 Garching, Germany}

\altaffiltext{4}{Harvard-Smithsonian Center for Astrophysics,
60 Garden St., Cambridge, MA 02138, USA}

\altaffiltext{5}{Department of Applied Mathematics and Theoretical 
Physics, The Queen's University of Belfast, Belfast BT7 1NN, UK}

\altaffiltext{6}{Astronomical Institute, Utrecht University, P.O. Box 
80000, 3507 TA Utrecht, the Netherlands}

\begin{abstract}

We find resolved interstellar O K, Ne K, and Fe L
absorption
spectra in the {\it Chandra} Low Energy Transmission Grating 
Spectrometer spectrum of the low mass X-ray binary X0614+091. 
We measure the column densities in O and Ne, and find 
direct spectroscopic constraints on the
chemical state of the interstellar O.
These measurements probably probe a low-density
line of sight through the Galaxy 
and we discuss the results in the context
of our knowledge of the properties of interstellar
matter in regions between the spiral arms.

\end{abstract}


\section{Introduction}

It has long been acknowledged that X-ray absorption spectroscopy
should be a powerful technique to address the physics and chemistry of
the interstellar medium (ISM). The X-ray band contains the K spectra
of all charge states of 
the abundant elements from C to Fe, so that the X-ray
spectrum simultaneously captures these elements regardless of their
ionization state. In addition, the X-ray 
absorption spectrum is sensitive to the physical and chemical
properties of the interstellar medium constituents. 
The wavelengths and shapes of
absorption lines and edges sensitively depend on whether the absorbing
atoms are free, or bound in molecules, and on
whether the atoms are the gas phase or locked in a solid. A first
demonstration of the potential of X-ray spectroscopy in this regard
was given by Schattenburg and Canizares (1986), who detected
absorption by interstellar O and Ne in the {\it Einstein}
Focal Plane Crystal Spectrometer observations of
the Crab nebula, and even
found weak evidence for the presence of both neutral and singly
ionized O along the line of sight. They also found a hint of the
expected narrow $1s-2p$ absorption line in atomic O. 

We obtained a high resolution spectrum of the bright, relatively 
unabsorbed low-mass X-ray binary X0614+091 with the Low Energy
Transmission Grating Spectrometer (LETGS) on the {\it Chandra} X-ray
observatory, hoping to resolve the
soft X-ray line emission that was detected with the {\it Einstein} SSS
(Christian, White, \& Swank 1994),
and the Solid
State Imaging Spectrometers on {\it ASCA} (White, Kallman, \& Angelini
1997).
To our disappointment, we find no evidence for narrow emission lines
in the current data, which we take to imply that the discrete emission
in this source is time variable. Instead, it turns out that the
spectrum of X0614+091 is ideal for interstellar absorption
spectroscopy: the source is bright, has a featureless intrinsic 
continuum, and happens to be behind a column density of interstellar
matter such that we have optimum contrast at the O K absorption
edge. We detect K shell absorption by interstellar
O and Ne, and L shell absorption by Fe.

\section{Data Analysis}

X0614+091 was observed on November 28, 1999, starting at 22:26:07 UT,
for 27,390 seconds. The LETGS has been described in Brinkman et al.
(1987, 1997), Predehl et al. (1997), and in Brinkman et al. (2000). 
Our observation used the Low Energy Transmission Grating in conjunction with
the High Resolution Camera (HRC-S) microchannel plate detector (Murray et
al. 1997). The LETGS has highest sensitivity in the 5--150 \AA\ (0.08--2.5
keV)
band, with an 
approximately constant wavelength resolution of $\approx 0.06$ \AA.
The data were processed by the standard {\it Chandra} X-ray Center
pipeline, and sorted into an image. The spectrum was extracted by
placing a rectangular mask along the spectral image and summing the
photons in the cross-dispersion direction. Background was determined
from rectangular strips parallel to the spectrum, and offset in the
cross-dispersion direction. Since we will only 
be using data at wavelengths less
than approximately 30 \AA, the use of rectangular masks is adequate.

The location of the centroid of the zero order image was determined,
which fixes the zero of the wavelength scale. Pixel numbers were
converted to wavelength using the calibration of the wavelength scale 
derived from the spectrum of Capella (Brinkman et al. 2000). This
wavelength scale is currently believed to be accurate to approximately
20 m\AA (status as of May 2000). 
The residual uncertainties appear to be mainly determined by
remaining
small systematic distortions of the spectral image by the detector and
possibly the effect of line blending.
The data were binned in 0.02 \AA\ bins, and the positive and
negative spectral orders were added to enhance the signal-to-noise
ratio, especially longward of 20 \AA. In the present paper, we will not be
using absolute fluxes, basing all quantitative conclusions on the shape of
the spectrum over narrow spectral ranges only. We therefore ignore the small
systematic errors in the absolute fluxes resulting from our reduction
procedure, that are 
associated with spatial variations in the detector
efficiency and background rate.

We removed the higher order contributions to the spectrum in the
following manner. Starting at the shortest wavelengths, for each
wavelength bin $[\lambda_n,\lambda_{n+1}]$, we compute the
number of counts in the second order by scaling with the ratio of the
grating efficiency in order $m=2$ to the first-order efficiency, at
$\lambda_n$. Note that the efficiencies of all other instrument components
divide out, with the exception of the detector flat field.
These second-order counts are then subtracted from the counts in the
interval $2 \cdot [\lambda_n,\lambda_{n+1}]$. The process is repeated
for orders up to $m=6$; going to higher $m$ does not affect the
spectrum below 30 \AA\ anymore, because the first order spectrum
decreases steeply shortward of 6 \AA. We distribute the 
higher orders counts to be subtracted from the spectrum  
uniformly over the bins in the wavelength range
$m \cdot [\lambda_n,\lambda_{n+1}]$.
This assumption concerning the wavelength distribution of the higher
order counts is justified for what is evidently a smooth continuum 
spectrum, and it avoids the large amplification of noise that would
result from choosing a narrower redistribution. It does artifically
degrade the resolution of the estimated higher order spectrum,
but since the incident spectrum is very smooth
we will ignore this inaccuracy here. 

Figure 1 shows the spectrum between 5 and 30 \AA,
background-subtracted and higher orders removed. The background
amounts to approximately 1.5--1.8 counts bin$^{-1}$ over the range
10--40 \AA. The cumulative higher order flux equals the first order flux at
approximately 26 \AA, and exceeds it at longer wavelengths.

As is evident, we
detect no significant line emission. Instead, a number of highly
significant absorption features are clearly visible in the spectrum. 
The absorption features appear in both the positive and negative
spectral orders, at mutually consistent wavelengths. 
The features
coincide with the K absorption edges in neutral Ne and O, and the
L$_{2,3}$ edges in neutral Fe. The O edge is much deeper than a
small edge in the instrument, and is clearly of interstellar origin.
In addition, we detect a narrow feature
at $\sim 23.5$ \AA, consistent with the wavelength of the expected
$1s-2p$ absorption line in neutral atomic O. All these features are
consistent with absorption expected from the ISM. 
In the following
section, we will discuss the interstellar absorption spectrum in detail.
The deep edge at 6 \AA\ is instrumental, and comprises the
M$_{\rm 4,5}$ edges in the Ir coating of the mirror, and the M$_{\rm 4,5}$
edges in the grating material, Au. In addition, we see the 
M$_{\rm 4,5}$ edges from Cs and I in the CsI coating of the detector. 
All relevant instrumental features are indicated in Figure 1.

Work on the calibration of the effective area of the LETGS is still in
progress. Currently, the effective area is not known accurately enough
across the entire 5--30 \AA\ band to attempt a quantitative
characterization of the overall spectrum. However, we note that a very
simple continuum model (a constant flux in units photons cm$^{-2}$ s$^{-1}$
\AA$^{-1}$, absorbed by an interstellar column density of neutral gas of 
$N_{\rm H} \sim 1.5-2.0 \times 10^{21}$ cm$^{-1}$) 
describes the data to within
$\approx$ 20\% across the 5--30 \AA\ band. This implies a total unabsorbed
flux
in our spectrum of 
$F(5-30~{\rm \AA}) \approx 1 \times 10^{-9}$ erg cm$^{-2}$ s$^{-1}$.
The spectral shape is very similar to that found in previous observations,
with {\it Einstein} (Christian, White \& Swank 1994) and
{\it Beppo}SAX (Piraino et al. 1999). The total unabsorbed flux is in the
middle of the range of fluxes 
observed with {\it Einstein} (when converted to the
0.5--20 keV band quoted by Christian, White, \& Swank [1994]), and is 
virtually equal to that observed with {\it Beppo}SAX (when converted to the
{\it Beppo}SAX 0.1--200 keV band).

\section{Absorption by Interstellar O, Ne, and Fe}

In Figure 2 we show the spectrum in the wavelength range centered on the O K
edge. The counting statistical fluctuations in this range of the spectrum
amount to approximately 8 counts
bin$^{-1}$ on average, and a cumulative higher order flux of approximately 
25 counts bin$^{-1}$ has been removed. A deep edge and a strong narrow
absorption line are readily visible, as well as perhaps
some additional structure. 

There is some structure in the spectrum due to absorption by O in the
UV-Ion Shield (UVIS) on the HRC (Murray et al. 1997). The UVIS
comprises a $\sim 800$ \AA\ layer of  Al on a 2750 \AA\ film of polyimide
in the central regions of the detector, and a thinner (300 \AA) layer of
Al at longer wavelengths (beyond $\sim 18$ \AA). The boundary between the
thick and thin Al filters is washed out over an approximately 3
\AA\ wide band due to the 
dithering of the spacecraft, and because the filter is mounted out of 
focus. The jump in transmission across this region is $\lax 10$ \%.
The detailed transmission characteristics of the UVIS in the region
of
the C, N, and O edges were determined from measurements of witness
samples undertaken at the BESSY synchrotron facility using a plane
grating monochromator with a resolving power
$\lambda/\Delta\lambda\sim 2000$---approximately twice that of the
LETGS in the same spectral regions.
We show the absorption expected from the UVIS explicitly in Figure
2.
A shallow, narrow absorption
feature appears at $\approx 23.3$ \AA, as well as an 'edge', which gradually
rolls over between 22.9 and 23.2 \AA, with a relative depth of
$\approx 25$ \% (see Figure 2).
The narrow feature is most likely due to absorption by O bound to N and C
in the polyimide, by the analogue of the $1s-2p$ transition, 
while the continuum
absorption is due to all O in the UVIS (polyimide and aluminum oxide).
Given the stability of the materials, 
the transmission curve is not expected to
have changed since the ground calibration. 

The deep narrow  resonance at $\approx 23.5$ \AA\ is undoubtedly due to 
the $1s-2p$ absorption line 
in interstellar monatomic neutral O. Table 1 lists measured and
calculated values for the wavelength of this transition. We measured the
wavelength from our data by fitting a narrow Lorentzian absorption line,
convolved with a Gaussian of width 0.06 \AA\ (FWHM) to represent the 
response of the LETGS (Brinkman et al. 2000). We find the centroid
wavelength to be $\lambda_0 = 23.52 \pm 0.02$ \AA, where the error
represents the 90\% confidence range for one parameter of interest. We take
the systematic uncertainty in the wavelength scale into account by adding 
0.02 \AA\ in quadrature to this uncertainty. As can be seen in Table 1,
our value is consistent with modern experiments (even if the published
experimental values do not agree with each other). 

McLaughlin \& Kirby (1998) calculated the detailed K-shell photoabsorption 
cross section for atomic O using the R-matrix technique. We use their data
to model the absorption by interstellar O. Figure 2 shows the absorption by
a column density of O atoms of $N_{\rm O}
= 8 \times 10^{17}$ cm$^{-2}$. The model
spectrum has been shifted by 0.051 \AA\ to match the measured centroid
wavelength of the resonance. The continuum model is a simple powerlaw, with
interstellar absorption by elements other than O modeled with the cross
sections given by Morrison \& McCammon (1983). The entire model spectrum is
multiplied with the effective area of the LETGS before convolution with the
instrument response.

The measured column density of O, $N_{\rm O}
\approx 8 \times 10^{17}$ cm$^{-2}$,
implies that the absorption line is saturated: the optical depth at line
center is $\tau_0 \approx 40$ (the line is not resolved by the
spectrometer). Unfortunately, that means that the line depth or equivalent
width are insensitive to the column density of atomic O. 
The depth and shape
of the K-edge are reasonably well represented by the model, but there
appears to be some additional absorption near the edge (22.7--23.0 \AA), and
there is marginal evidence (3--4$\sigma$) for a 
narrow absorption feature at $\approx 23.4$ \AA, which could be
due to absorption by ionized O, or O in molecules or dust.

Wavelengths for the narrow resonance in CO, various iron oxides, and in
singly ionized O, are listed in Table 2. As expected, these wavelengths are
offset from the wavelength of the resonance in atomic O by a few parts in
500, roughly the ratio of the $n=2$ binding energy per valence electron to
the $1s$ binding energy. We also list the $1s$ binding energy for neutral
and singly ionized O. 
There is no obvious absorption edge due to O$^+$ shortward of the neutral O
ionization limit, and we place a rough upper limit of $\tau_{\rm O^+} \lax
0.2$ on the optical depth in the O$^+$ continuum. Assuming that the K-shell
photoionization cross sections for O and O$^+$ are approximately equal, this
implies $N_{\rm O^+} \lax 6 \times 10^{17}$ cm$^{-2}$. 
It should be noted that the
wavelengths for the transitions in O$^+$ listed in Table 2 were obtained
from approximate atomic structure calculations, and they could be in error
by up to of order 1\%, or 0.25 \AA. 

The wavelengths listed in Table 2 indicate that we can exclude significant
absorption by the abundant molecular species CO. This is illustrated in
Figure 3, where we show the absorption spectrum due to a column density 
$N_{\rm CO} = 2 \times 10^{17}$ cm$^{-2}$, in addition to an atomic column
density of $N_{\rm O} = 6 \times 10^{17}$ cm$^{-2}$. The cross section for
CO was taken from Barrus et al. (1979). This combination of column densities
was chosen for illustrative purposes only; it produces almost the
same continuum absorption as 
$N_{\rm O} = 8 \times 10^{17}$ cm$^{-2}$ of
pure atomic absorption, but produces a strong CO
resonance feature near 23.2 \AA, which is not observed. This absence
allows us to place a rough upper limit on the column density
of O in CO, of approximately $\lax 5 \times 10^{16}$ CO molecules cm$^{-2}$,
as judged by eye. The statistical quality of the spectrum probably does not
warrant a more complicated quantitative analysis in this respect.

Interestingly, the feature near 23.4 \AA\ coincides with measured
wavelengths for the narrow resonance in iron oxides. Wu et al. (1997) give
the relative shapes of the O K-absorption spectra in various iron oxides. In
addition to the narrow feature at 23.4 \AA, the oxides also have deep,
broad continuum absorption, with maximum absorption in the range 22.8--23.0
\AA. We estimate the depth of the narrow feature to be roughly 
$\tau_0 \sim 0.5-1.0$. Again, the 
noise in the data has large enough amplitude that
a full quantitative statistical analysis is not warranted. 
Unfortunately, we are not aware of any quantitative measurements or
calculations for the absolute magnitude of the photoabsorption cross
section in iron oxide. But if we assume that the total absorption cross
section, integrated over the resonance, has the same value as for atomic O,
and that the line is unresolved,
then this line depth correponds to an equivalent column density of O atoms
of $N_{\rm O} \sim 1-2 \times 10^{16}$ cm$^{-2}$. Given the cosmic abundance
ratio of Fe to O, $n({\rm Fe})/n({\rm O}) = 5.5 \times 10^{-2}$ (Anders \&
Grevesse 1989),
this result would imply that a major fraction (up to half)
of the Fe could be in oxides.

Figure 4 shows the wavelength range between 13 and 19 \AA\ on an expanded
scale. In this range are the K edge of Ne, and the L edges of Fe. 
Background is negligibly small compared to the first order flux in this
region of the spectrum. The average level of the higher order radiation
amounts to 20--30 counts bin$^{-1}$.
We have indicated the locations of the Fe L$_{1,2,3}$ edges, taken from 
Fuggle \& M{\aa}rtensson (1980, quoted in Williams 1986; these are
binding energies relative to the Fermi level). The L$_{2,3}$ edges 
are clearly visible in the spectrum.
The L$_1$ ($2s$) edge 
is expected to be much weaker than the $2p$ edges, and is indeed not
detected. 
The Fe L edges are well-separated from the
instrumental Cs M$_{4,5}$ edges, which are expected to appear in emission
(they appear in emission in the featureless
continuum spectrum of the quasar 3C273, see Kaastra et al. 2000). The Fe
absorption could be due to metallic iron, iron in molecules, or both.
Unfortunately, we have been unable to locate detailed measured absorption cross
sections for metallic Fe in the literature.

Crocombette et al. (1995) give experimental and theoretical 
$2p$ absorption
spectra for iron oxides (FeO, Fe$_2$O$_3$, Fe$_3$O$_4$), which we have scaled
and superimposed on the data in Fig. 4. The general shape of the oxide
absorption spectrum 
is similar to that observed here, except that the measured wavelengths are
significantly longer than those observed in our spectrum (by approximately
0.2 \AA, for all three oxides). We note that a comparison of the positive and
negative spectral orders in this limited region seems to indicate a mismatch
of about 0.08 \AA. The Fe $2p$ features as they appear in Figure 4 are at the
average of the positive and negative orders.
We caution that this mismatch is larger than any of the wavelength
residuals observed in the emission lines in the spectrum of Capella, but
it
may just be a statistical fluctuation. Taken at face value,
the wavelength offset between
the Fe oxide absorption spectrum and the Fe $2p$ absorption spectrum observed
in our data indicates that (most of) the interstellar absorption is not due to
Fe oxides, and in fact is consistent with the expected wavelengths for
absorption in metallic Fe ! 
This is not consistent with our tentative conclusion from the analysis of
the O spectrum that a major fraction of the Fe along this line of sight
may be in oxides. A possible explanation is that there is a systematic error
on the wavelength scale used by Crocombette et al. (1995), but we are
unfortunately in no position to judge the likelihood of this possibility,
except to note that the wavelength discrepancy appears large by laboratory
experimental standards.
In view of the possible small distortion of our wavelength
scale, we choose to leave this issue open for now.

Finally, analysis
of the Ne K edge is relatively straightforward. We use the cross section
given by Verner et al. (1993) to derive a column density of neutral Ne of 
$N_{\rm Ne} = 1.0 \times 10^{18}$ cm$^{-2}$, with an estimated uncertainty of
about 20\%. There is some confusion about the exact ionization energy.
Verner et al. (1993) give a wavelength of 14.45 \AA\ for the edge,
while Cardona \& Ley (1978, quoted in Williams 1986) give 14.25 \AA. In
our data, the edge actually appears approximately halfway between these two
values. There also appears to be some additional structure, a narrow
absorption feature at 14.45 \AA. This cannot be due to a narrow $1s-2p$
resonance in neutral Ne, because neutral Ne has a closed $n=2$ shell.
A narrow $1s-2p$ absorption line in singly ionized Ne is
technically possible. Kaastra \& Mewe (1993) list 14.38 \AA\ for the
wavelength of the $1s-2p$ transition in Ne$^+$, which has a systematic
uncertainty of order 1\%. 
However, we do not observe a Ne$^+$ $1s$ ionization edge, 
for which Verner et al. (1993) list a wavelength of 14.04 \AA.
Alternatively, the feature could be the $1s-3p$ transition in neutral
Ne (compare the K absorption spectrum of neutral Ar in 
Parratt 1939).
Unfortunately, we are not aware of
any experimentally determined wavelengths for the K spectrum of Ne$^{+0}$ or
Ne$^{+1}$. In either
case, the total depth of the K edge is a reasonable measure of the total
column density of Ne along the line of sight. 

\section{Discussion}

We have measured the X-ray absorption spectrum of the interstellar medium
along the line of sight to the Low-Mass X-ray Binary X0614+091. We detect
the K-absorption spectra of Ne and O, and the L absorption spectrum of Fe. 
The spectra are of sufficiently high resolution and sensitivity that they
are sensitive in principle to the physical and chemical state of the ISM.
However, our ability to extract the full measure of quantitative information
from our data is limited in important respects by the current lack of accurate
laboratory measurements or reliable calculations of transition wavelengths
and absorption cross sections for the various ions and
molecular species of the abundant elements.

The following conclusions are robust, however. We measure a total
interstellar oxygen column density of $N_{\rm O} = 8 \times 10^{17}$
cm$^{-2}$, a fraction of which may be in
ionized O or O bound in molecular gas, oxides, or dust. We find that this
bound or ionized 
O cannot all be in CO. We set an upper limit on the fraction of O
in CO of $N_{\rm CO}/N_{\rm O} \lax 0.06$. 
Instead, the shape of the spectrum appears
consistent with the presence of a significant amount of iron oxide, but this
conclusion is not unique, given the limited amount of laboratory comparison
spectroscopy. If the narrow resonance feature at $\sim 23.3$ \AA\ is indeed
due to iron oxide, we estimate an equivalent column density of O in iron
oxide of $N_{\rm O} \sim 1-2 \times 10^{16}$ cm$^{-2}$, {\it i.e.} a
fraction $1-3 \times 10^{-2}$ of the atomic O along the line of sight, which
would be bound to a major fraction of the available iron.

We measure a total Ne column density of $N_{\rm Ne} = 1.0 \times 10^{18}$
cm$^{-2}$, which may contain an uncertain fraction of singly ionized Ne.
This implies an abundance ratio O/Ne $\approx 0.8$, whereas the solar
ratio is 6.0 (Anders \& Grevesse 1989). 
The optical depth in the O edge may have been underestimated if we did not
subtract enough higher order flux at the O edge. While there are some
uncertainties in the calibration of the diffraction efficiencies in the
various grating orders, and while we have neglected spatial variations in the
detector efficiency, it is unlikely that we have
underestimated the depth of the O edge by a factor 6. 
The discrepancy is also far too large to be ascribed to uncertainties in the
photoabsorption cross sections of either O or Ne, which typically run 
at 20--30 \% or less.

A possible
astrophysical explanation for the relative shallowness of the O edge
would be that a major fraction of the interstellar O along this line of
sight is in fact locked up in optically thick dust grains. However, this
appears implausible, because it would require the dust grains to have an
average size of order $\sim 1 \mu$m or larger, 
whereas the characteristic size of
galactic dust grains is closer to 0.1 $\mu$m, as derived directly from
X-ray scattering by interstellar dust (Mauche \& Gorenstein 1986;
Predehl \& Klose 1996). The only other published
X-ray absorption spectrum around the
O K edge in fact indicates that the dust grains must be transparent to 
soft X-rays (Schattenburg \& Canizares 1986).

In fact, it appears that Ne may be overabundant, instead of O being
underabundant. If we convert the measured column densities to equivalent H
column densities using cosmic abundances, we find $N_{\rm H} \sim 9 \times
10^{20}$ cm$^{-2}$ from the O column density, and $N_{\rm H} \sim 8 \times
10^{21}$ cm$^{-2}$ from the Ne column density. The latter figure is a factor
4--6 larger than the H column density we infer from the overal shape of the
spectrum (see section 2), 
whereas the H column density implied by the O column is within a
factor of about 2 of that estimate. 

In principle, this apparent overabundance of Ne could be associated
with absorbing gas in the binary (a large overabundance of Ne in the
general ISM appears very unlikely), or with gas blown out of the binary
system. In this context, it is perhaps interesting to note that at least one
other low-mass X-ray binary, 4U1627-67, has also shown a large apparent
overabundance of Ne. In the case of 4U1627-67, this inference is based on
the presence of a very bright Ne X Ly$\alpha$ emission line in the X-ray
emission spectrum observed with {\it ASCA} (Angelini et al. 1995).
For now, we defer interpretation of the apparent overabundance of Ne in the
absorption spectrum of X0614+091
until after more
astrophysical X-ray absorption spectra have become available.

For a Low-Mass X-ray Binary, X0614+091 is located in a somewhat unusual 
direction: $l_{\rm II} = 200.9$ deg, $b_{\rm II} = -3.4$ deg, 
{\it i.e.} more or
less in the direction of the galactic anticenter, a few degrees below the
galactic plane. The distance to X0614+091 has been constrained to $d \lax 3$
kpc, from the requirement that the inferred peak luminosity during an X-ray
burst not exceed the Eddington limit (Brandt et al. 1992). 
From the reddening of the optical
counterpart, Machin et al. (1990) conclude that the source is at least $\sim
1.5$ kpc away, so that it is located either just at the near edge of the
Perseus arm, or inside or behind it (see the recent spiral pattern map in
Taylor \& Cordes 1993, their Figure 1). Given the fairly low column density of 
$N_{\rm H} = 1-2 \times 10^{21}$ cm $^{-2}$, 
most of the absorption is likely to occur in
the low-density environment between the spiral arms, however. The upper
limit to the CO column density is not suprising in that case. Note that this
region of the sky is just outside the coverage in the deep CO survey of the
third galactic
quadrant (May et al. 1993). 
If the source is indeed located on the near side of the Perseus arm, the
possible detection of iron oxides, signaling the presence of a
non-negligible amount of dust in the interarm region, may be interesting. 

\acknowledgements

We gratefully acknowledge conversations with Jacqueline van
Gorkom, and a very thorough and helpful reading by the anonymous referee.
FP acknowledges support from NASA under contract NAS5-31429.

\newpage

\begin{deluxetable}{lll}

\tablewidth{0pt}
\tablecaption{Wavelengths for Transitions in Atomic Oxygen}

\tablehead{
\colhead{reference} & \colhead{$1s-2p$} &
\colhead{$1s-\infty$}  
}

\startdata

Experiment, Krause (1994) &
23.489 $\pm$ 0.0044 \AA & 22.775 $\pm$ 0.013 \AA \\
Experiment, Stolte et al. (1997) &
23.536 $\pm$ 0.0018 & 22.790 $\pm$ 0.002 \\
Experiment, This Work &
23.52 $\pm$ 0.03 & ...\\
Theory, McLaughlin \& Kirby (1998)\tablenotemark{1} &
23.467  & (22.775) \\
Theory, McLaughlin \& Kirby (1998)\tablenotemark{2} &
23.472  & (22.790) \\

\enddata

\tablenotetext{1}{R-matrix technique, 
with the energy of the series limit adjusted to 544.40 eV}
\tablenotetext{2}{R-matrix technique, 
with the energy of the series limit adjusted to 544.03 eV}

\end{deluxetable}

\begin{deluxetable}{lccl}

\tablewidth{0pt}
\tablecaption{Wavelengths for Transitions in Atomic and Molecular
Oxygen}

\tablehead{
\colhead{transition} & \colhead{wavelength} &
\colhead{width\tablenotemark{1}} & 
\colhead{reference} \\
\colhead{} & \colhead{(\AA)} & \colhead{(\AA)} & \colhead{}
}

\startdata

$1s-2p$, atomic O      & 23.467 & 0.0082 & McLaughlin \& Kirby (1998) \\
$1s-2p$, O$^+$         & 23.40  & ...    & Kaastra \& Mewe (1993) \\
$1s-2p$, CO          & 23.21  & 0.07   & Barrus et al. (1979) \\
$1s-2p$, FeO         & 23.31  & $\sim 0.1$ & Wu et al. (1997) \\
$1s-2p$, Fe$_2$O$_3$ & 23.36, 23.31  & $\sim 0.1$ & Wu et al. (1997) \\
$1s-2p$, Fe$_3$O$_4$ & 23.29  & $\sim 0.1$ & Wu et al. (1997) \\
$1s-\infty$, atomic O  & 22.78  & ---   & McLaughlin \& Kirby (1998) \\
$1s-\infty$, O$^+$     & 22.21  & ---   & Verner et al. (1993) \\

\enddata

\tablenotetext{1}{Full Width at Half-Maximum}

\end{deluxetable}

\newpage

\noindent
Figure Captions:

\figcaption{
The full 3.5--30 \AA\ spectrum of X0614+091, background-- and
higher-order subtracted, binned in 0.02 \AA\ bins. Both positive and
negative spectral orders have been summed. The wavelengths of the
interstellar Ne K, Fe L, and O K edges have been indicated, as well as the
wavelength of the narrow $1s-2p$ absorption line in atomic O. The location
of known instrumental features (Au, Ir, Al,
Cs, and I) has also been indicated. The weak feature at 25 \AA\ could
be due to an N-shell resonance in Ir.
}

\figcaption{
The range 21--25 \AA\ in the spectrum of X0614+091. The upper solid
line is a simple continuum model, multiplied with the spectrometer effective
area (see text). A shallow O K edge is visible at 22.9--23.2 \AA, as well as
a discrete feature at 23.3 \AA, due to the analogue of the $1s-2p$
transition in O bound to C and/or N. Both these features are due to the beam
filter in the instrument. The lower solid line is the same model, but with
absorption by a column density of $8 \times 10^{17}$ neutral
O atoms cm$^{-2}$ added. The location of the $1s-2p$ analogue in ionized O,
CO, and iron oxides has been indicated.
}

\figcaption{
As Figure 2, but with a column density of $2 \times 10^{17}$ O
atoms cm$^{-2}$ replaced by CO (lower solid line). A deep discrete feature
appears at $\sim 23.2$ \AA, the analogue of the $1s-2p$ transition in O
bound to C, which is not seen in the interstellar absorption spectrum of
X0614+091. The upper dotted line shows the absorption due to just the
neutral O column density of $6 \times 10^{17}$ atoms cm$^{-2}$.
}

\figcaption{
The spectral range 13--19 \AA, which contains the Ne K and Fe L
edges. The solid lines indicate approximate fits to the absorption spectra
(see text). The location of the model Ne K absorption edge has been set at 
the value given by Cardona \& Ley (1975). A narrow absorption feature may be
present at $\approx 14.5$ \AA. 
}

\vfill\eject

\centerline{\null}
\vskip7.5truein
\includegraphics{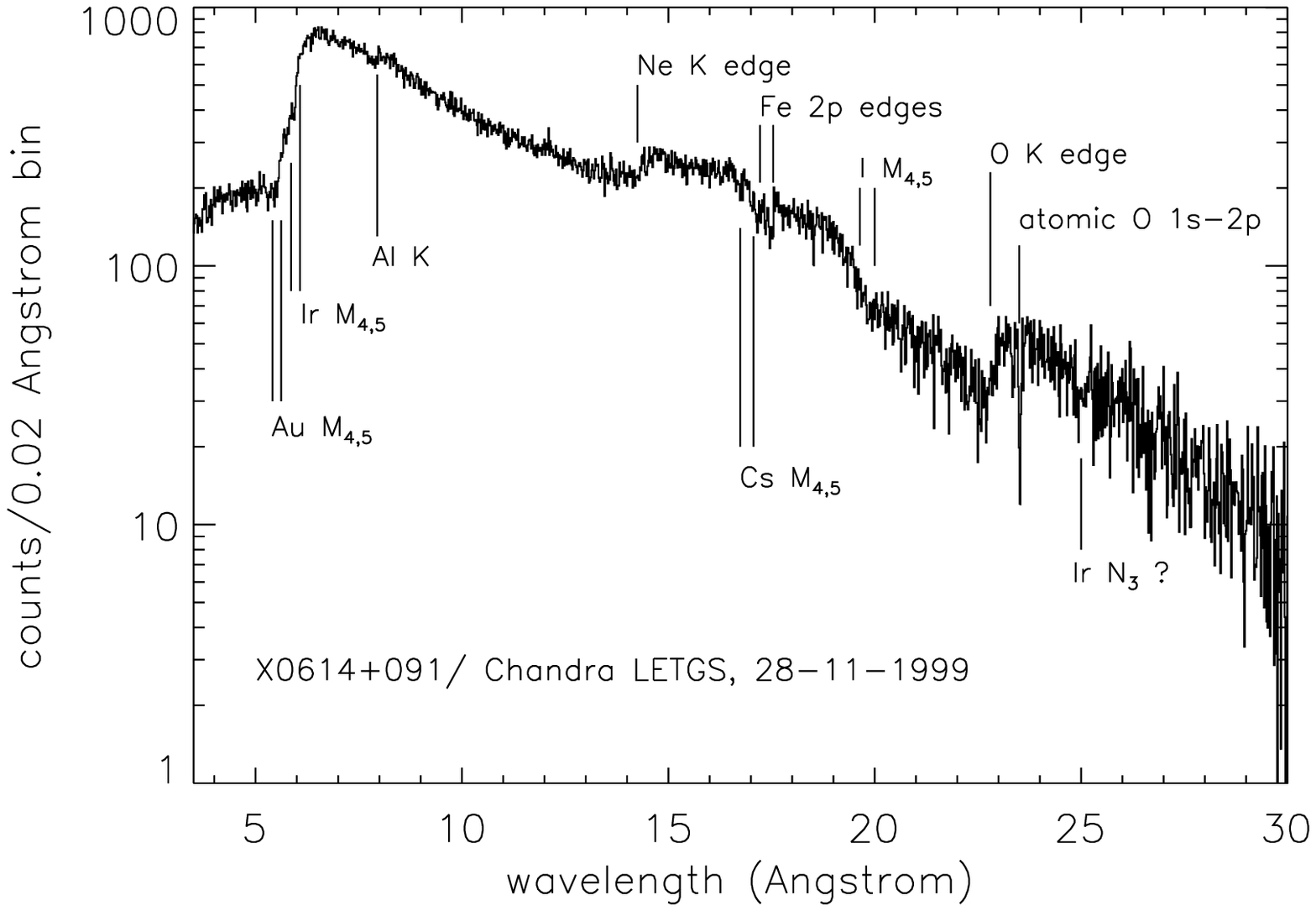}

\vfill\eject

\centerline{\null}
\vskip7.5truein
\includegraphics{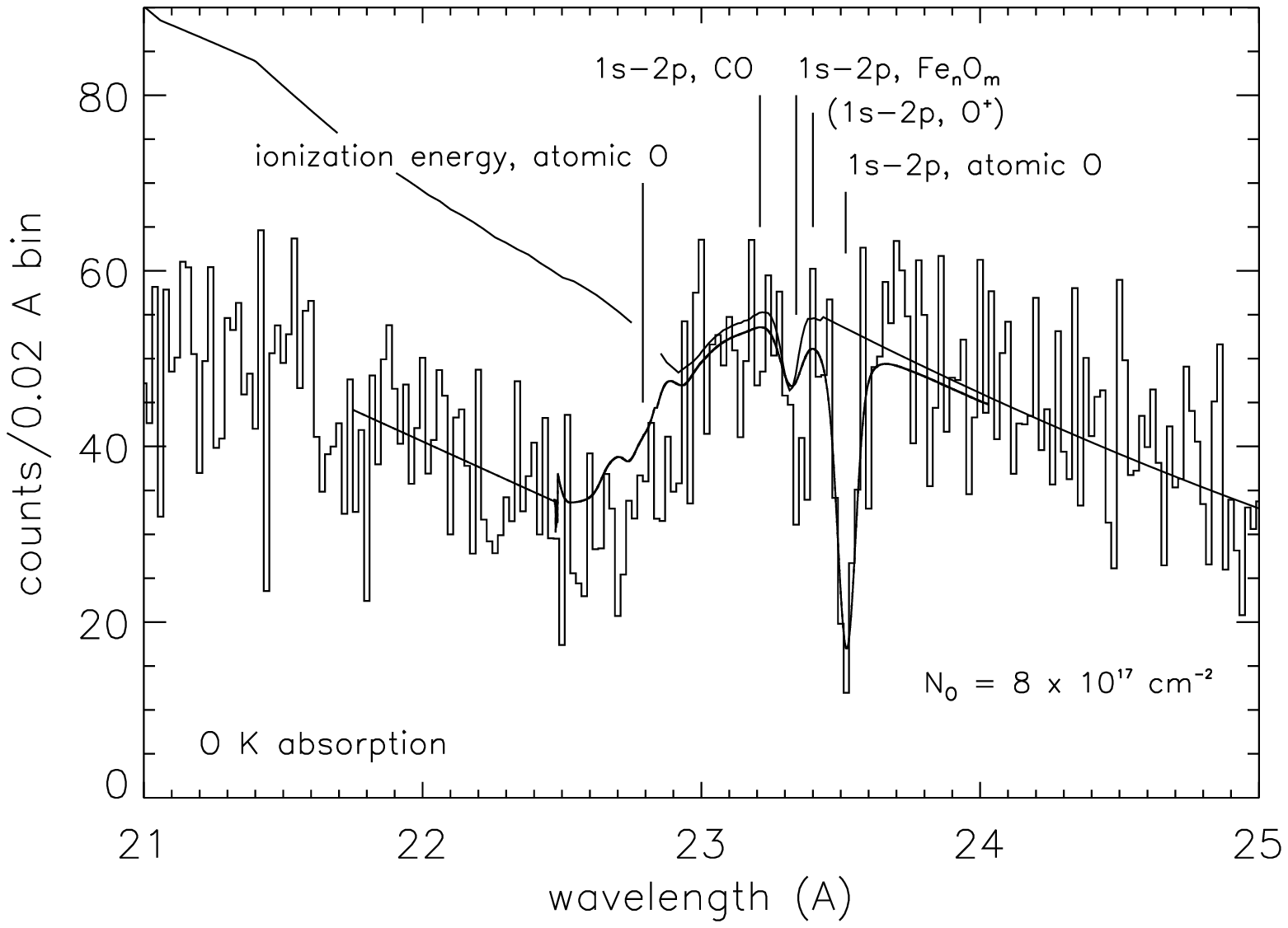}

\vfill\eject

\centerline{\null}
\vskip7.5truein
\includegraphics{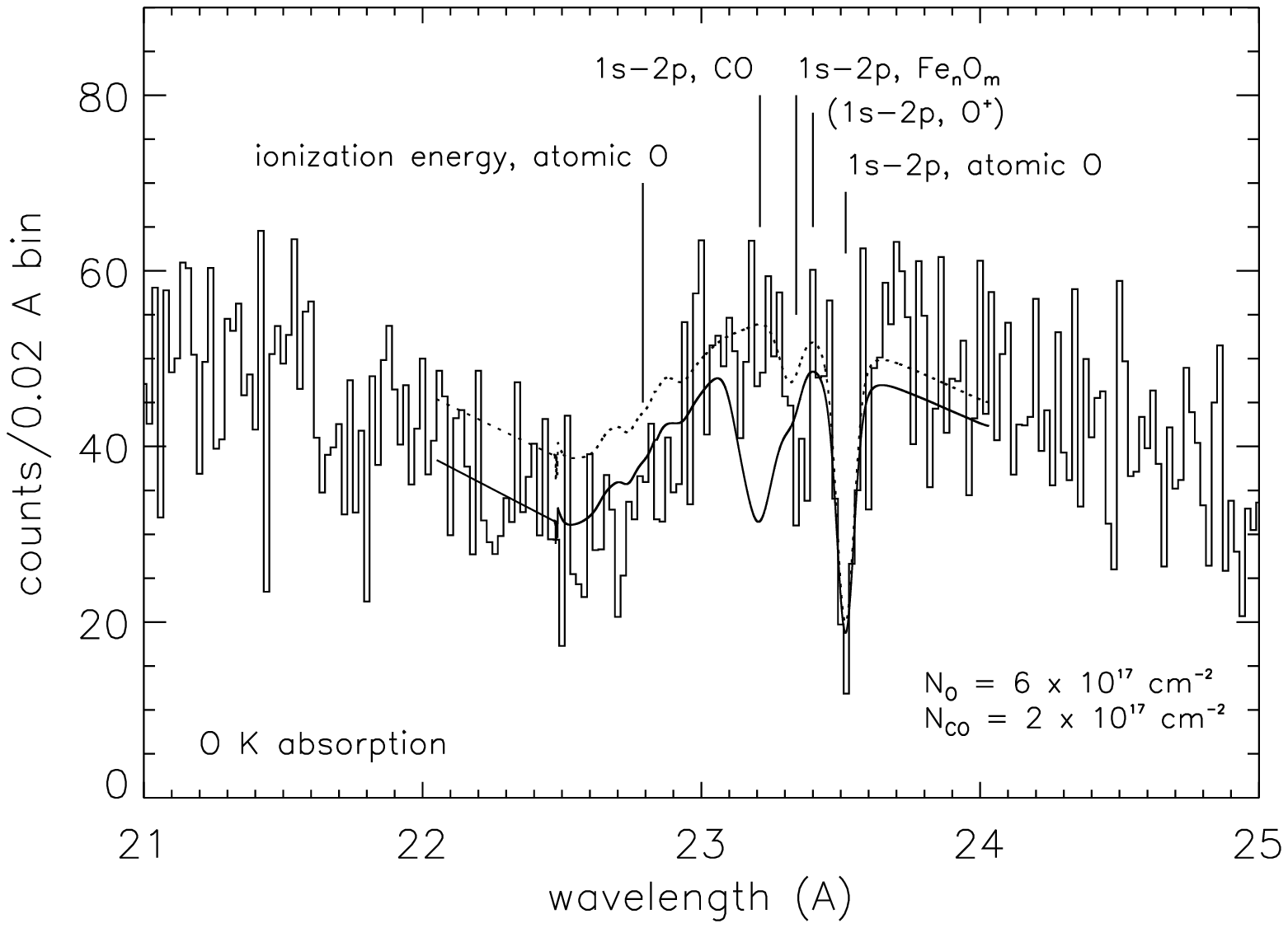}

\vfill\eject

\centerline{\null}
\vskip7.5truein
\includegraphics{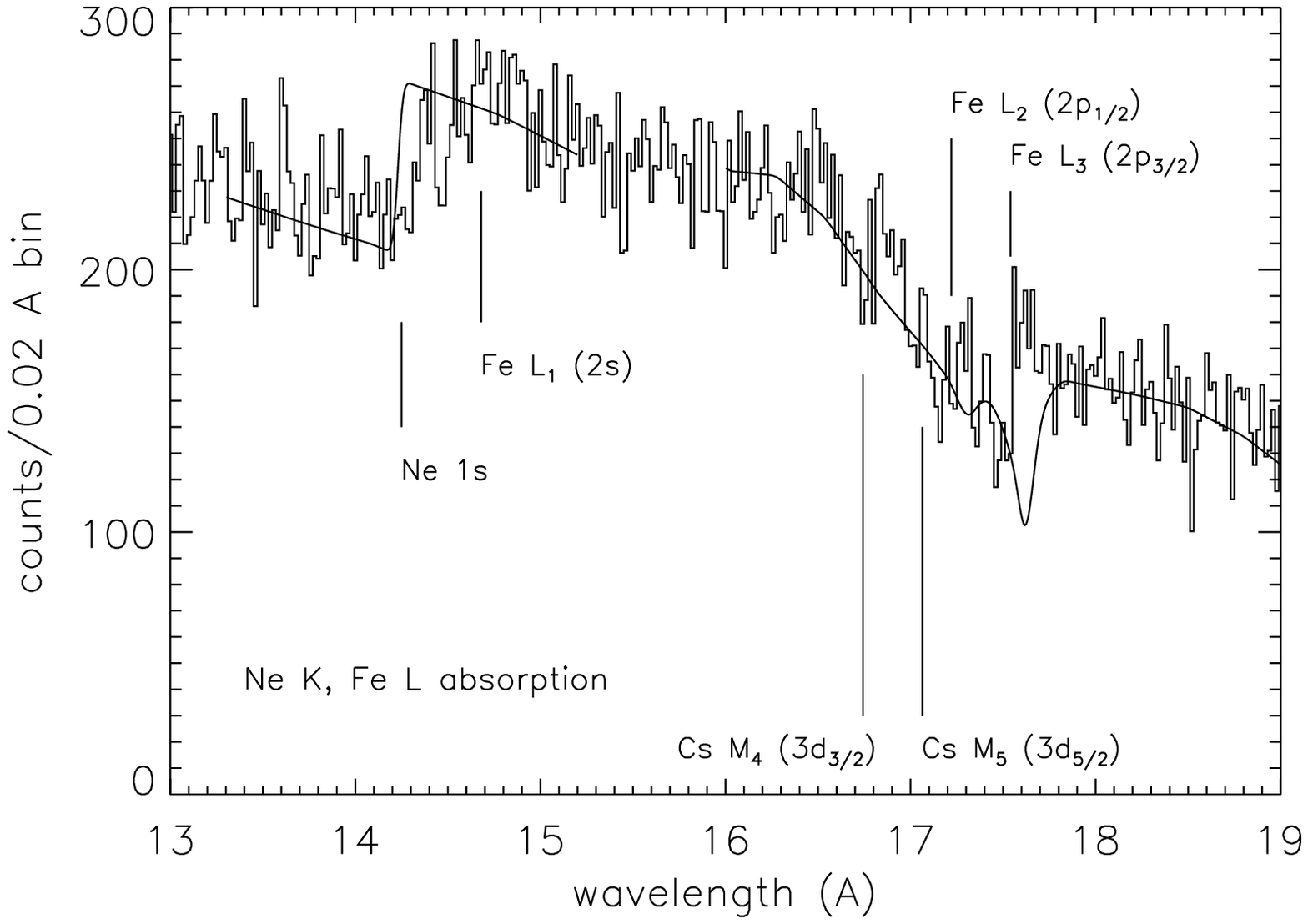}

\end{document}